\documentclass[conference]{IEEEtran}
\usepackage{blindtext, graphicx}
\usepackage{hyperref}
\usepackage{listings}
\usepackage{algorithm}
\usepackage{algpseudocode}
\usepackage[style=ieee]{biblatex} 
\begin{document}

\title{GPU Acceleration of Sparse Neural Networks}

\author{\IEEEauthorblockN{Aavaas Gajurel}
\IEEEauthorblockA{Department of Computer Science\\and Engineering\\
University of Nevada Reno\\
Reno, Nevada\\
Email: avs@nevada.unr.edu}
\and
\IEEEauthorblockN{Sushil J Louis}
\IEEEauthorblockA{Department of Computer Science\\and Engineering\\
University of Nevada Reno\\
Reno, Nevada\\
Email: sushil@cse.unr.edu}
\and
\IEEEauthorblockN{Frederick C Harris,Jr.}
\IEEEauthorblockA{Department of Computer Science\\and Engineering\\
University of Nevada Reno\\
Reno, Nevada\\
Email: fred.harris@cse.unr.edu}}

\maketitle

\begin{abstract}
In this paper, we use graphics processing units(GPU) to accelerate sparse and arbitrary structured neural networks. Sparse networks have nodes in the network that are not fully connected with nodes in preceding and following layers, and arbitrary structure neural networks have different number of nodes in each layers. Sparse Neural networks with arbitrary structures are generally created in the processes like neural network pruning and evolutionary machine learning strategies. We show that we can gain significant speedup for full activation of such neural networks using graphical processing units. We do a prepossessing step to determine dependency groups for all the nodes in a network, and use that information to guide the progression of activation in the neural network. Then we compute activation for each nodes in its own separate thread in the GPU, which allows for massive parallelization. We use CUDA framework to implement our approach and compare the results of sequential and GPU implementations. Our results show that the activation of sparse neural networks lends very well to GPU acceleration and can help speed up machine learning strategies which generate such networks or other processes that have similar structure. 

\end{abstract}


\section{Introduction}

Artificial neural networks, first proposed by Mcculoch and Pitts in 1943 \cite{mcculloch1943logical}, are universal function approximators loosely based on biological neural networks. Neural networks with back propagation \cite{hecht1992theory} is a robust method in machine learning. Artificial neural networks(ANN) have interconnected nodes that are separated into three types - inputs, outputs and hidden nodes. Inputs nodes are sensor nodes that take in values from outside the system, output nodes are the nodes that produce the answers from the network and hidden nodes are the nodes which lie in the information propagation path of the neural network. Each node is activated based on the nodes from which it has incoming connections, and the activation is calculated by weighting all the incoming node values with corresponding connection weight and summing all the values. The sum is then thresholded for the final activation. Generally the sum is passed through sigmoid function to constrain it within $-1$ and $+1$ values. 
$$sigmoid(x) = (1 / (1 + e^{-4.97*x}))$$

\begin{figure}[h]
\centering
\includegraphics[width=0.4\textwidth]{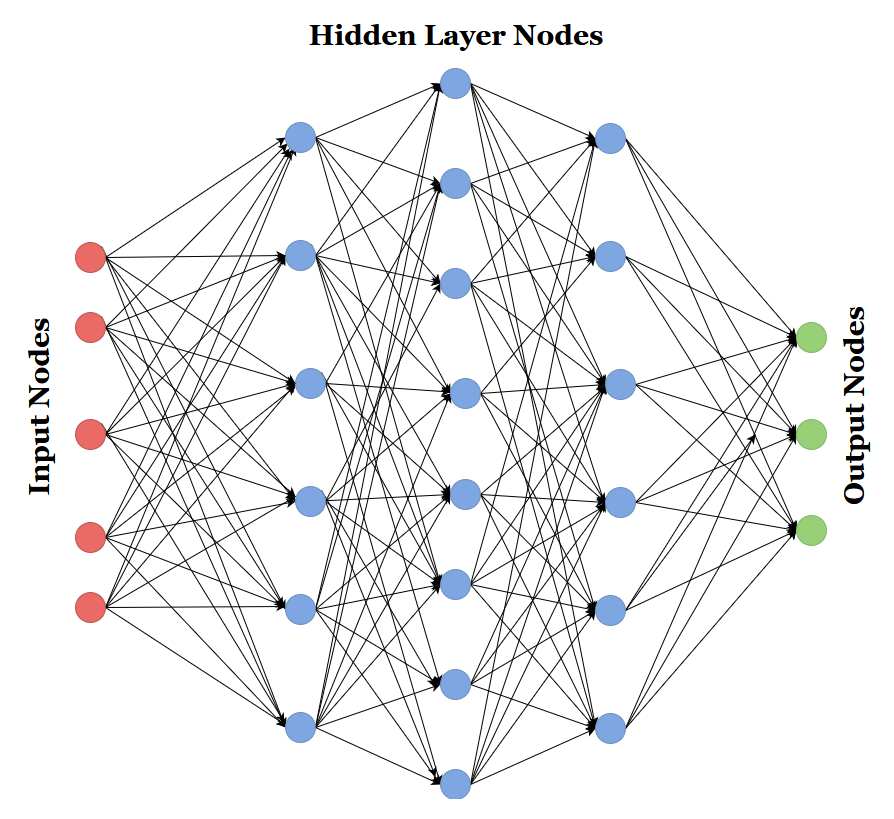}
\caption{Representation of conventional feed forward neural network with input, hidden and output layers}
\label{fig:ffnetwork}
\end{figure}

Conventionally, neural networks have structure which consists of nodes cleanly segmented into layers as shown in Figure \ref{fig:ffnetwork} with incoming nodes shown in red and outgoing nodes shown in green. Nodes in each layer are not connected with each other and are fully connected with the nodes in preceding layer and the following layer. This structure lends very well to vectorization where each layer is represented by matrices of weights and the propagation and activation can be calculated with the multiplication of the layer matrices. This property is utilized for GPU acceleration of such neural network. As GPUs are well suited for large in matrix multiplications, such neural networks have seen large speedups, even for huge networks \cite{fatahalian2004understanding}. 

\begin{figure}[h]
\centering
\includegraphics[width=0.4\textwidth]{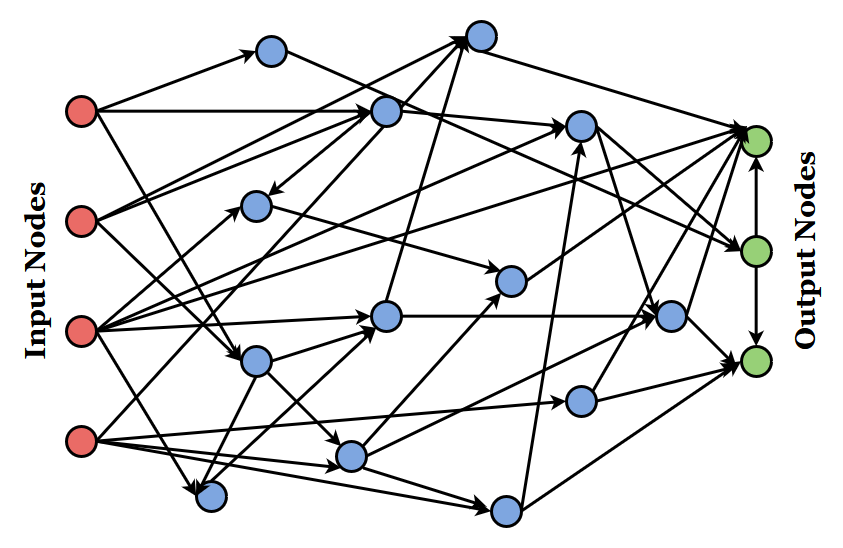}
\caption{Representation of neural network with arbitrary structure}
\label{fig:neatnetwork}
\end{figure}

This paper is concerned with the activation of sparse and arbitrary structured neural networks. Neurons in sparse neural networks do not have full connection with nodes from preceding layer and following layer. Neural networks have arbitrary structure when nodes from sparse networks are also pruned. In such case, such networks cannot be cleanly separated into layers, \textit{i.e.} they are not fully connected and can have incoming and outgoing links to any node in the graph as shown in Figure \ref{fig:neatnetwork}. Sparse networks are a subset of arbitrary structured neural networks(ASNN)s and are generated by neural network pruning algorithms \cite{lecun1990optimal} \cite{hassibi1993second}. ASNNs can be generated by pruning both connections and nodes from fully trained dense networks. They are also created by rule based network structure generators, and some of which are applied in machine learning to generate networks best fit for a given problem. Neural Evolution of Augmenting Topologies (NEAT) \cite{neatpaper} with direct encoding, HyperNEAT \cite{stanley2009hypercube} which uses generative encoding, and grammar based substitution and bi-directional growth encoding\cite{floreano2008neuroevolution} are some of the few processes which can generate ASNNs. With the start of application of neuro evolution to deep neural networks \cite{miikkulainen2017evolving}, full propagation of these network will take significant portion of compute time of total program running time. GPU acceleration methodologies used for conventional NN will not work with ASNNs, thus, if we can use GPU to accelerate arbitrary neural networks, we can have considerable gains in  speed and both memory and power efficiency.

Graphics processing units were originally developed as a co processor alongside the main CPU, to offload the processing of graphics related tasks which are are massively parallelizable. With co-accelerated advancements in games and GPU hardware, the GPU architecture reached a point where they could be used for general purpose computation. Initially, problems for general purpose computation in GPU(GPGPU) had to be converted into OPENGL shader programs utilizing non standard methods ~\cite{ino2006gpgpu} but with eager support from GPU manufactures, platforms for general purpose computation on GPU were created. Compute Unified Device Architecture (CUDA) ~\cite{kirk2007nvidia} and OpenCL ~\cite{stone2010opencl} being two prominent examples. With vendor support and capable frameworks, GPUs have become a useful tool for massive data parallel problems.

This new found power of general purpose computation on GPUs have been extensively utilized in neural network development and research. One of the reason for deep learning growth has been attributed to advancement in GPU capability \cite{strigl2010performance}. There are numerous GPU libraries supporting neural network acceleration \cite{nvidia2008cublas} and frameworks around them \cite{abadi2016tensorflow}. But GPU acceleration on arbitrary structure neural networks is still lacking; which we explore in this paper. For GPU acceleration of ASNN, we first perform a pre processing step that segregates all the nodes into dependency hierarchy which we call layers. Then we use thread parallalization available with CUDA interface to compute all the nodes in that layer at the same time. With our methodology, we have shown that we can get significant speedup with the use of GPUs and the speedup gets better with respect to increase in neural network connections and depth of the network. 

Remainder of this paper is organized as follows. Section ~\ref{RelatedWork} describes previous approaches related to our current work, section ~\ref{methodology} describes the neural network representation, sequential and GPU activation methodology and experimental setup. In Section ~\ref{results}, we describe our results and compare the timing and speedup between two strategies. lastly in section ~\ref{conclusion}, we draw our conclusions and explain possible future directions.

\section{Related Work} \label{RelatedWork}

GPUs have been used for general purpose computation from before the time they provided formal support for it. Before APIs like CUDA and OPENCL were offered by the GPU vendors, researchers utilised OPENGL shader languages to coerce the GPU into performing non graphical processing \cite{ino2006gpgpu}\cite{goddeke2005gpgpu}. The application of GPUs for accelerating data parallel tasks has only increased after the introduction of the supported APIs. NVIDIA maintains a host of different libraries targeted to various application domains like deepLearning, signal processing, linear algebra and others \cite{cudalibraries}.

The power of GPU have been eagerly utilized in the machine learning field as machine learning requires crunching through big numerical calculations and large number of iterations for making sense out of big datasets \cite{steinkraus2005using}. GPUs have been applied to Neural networks, and have been especially useful with the advent of deep learning, which uses neural networks with large number of neurons and hidden layers \cite{goodfellow2016deep}. Previous work on implementation of neural networks in GPUs have started before the introduction of CUDA, where the authors utilized texture processing pipeline of the GPU to accelerate Multi layer perceptron and self organizing maps with significant speedup \cite{luo2005artificial}. Scherer et. al. have shown in \cite{scherer2010accelerating} that GPU can have gains of up to two orders of magnitude for convolution neural networks. Cheltur et. al. have also shown that convolutional neural networks can be efficiently computed in GPUs with data framing in a form of matrix and performing matrix multiplication to compute the network \cite{chetlur2014cudnn}. Coates et. al. have shown that many consumer grade GPUS in separate machines can be used for acceleration of convolutional neural networks by using CUDA, and using openMPI for multi GPU coordination \cite{coates2013deep}. Zhang et. al. have also looked at accelerating sparse neural networks with custom hardware accelerators \cite{zhang2016cambricon}.

Other types of neural networks have also seen good results from GPU implementation. Nageswaran et. al. have implemented a configurable simulation environment for the efficient simulation of large scale spiking neural networks on GPU \cite{nageswaran2009configurable}. Juang et. al. have also shown significant speedup on fuzzy neural networks with high dimensional inputs by using parallel processing on GPUS \cite{juang2011speedup}, and GPUS have also been able to reduce recurrent neural networks training time by a factor of 32 \cite{chen2014efficient}. 

Neural network in general form are also graph structures, and there have also been numerous research on graph processing on GPUs. Luo et. al. showed that the speedup of upto 10x could be achieved with GPU implementation for breath first search \cite{luo2010effective}. Harish and Narayan give implementation of various graph processing algorithms on GPU in \cite{harish2007accelerating} and note that in some cases sequential approach does not transfer well to the GPU approach. In this paper, we are looking into networks with non uniform structure and have to perform a pre processing step on that structure to segregate the nodes where we have to apply graph processing approaches.

\section{Methodology} \label{methodology}

\subsection{CUDA}

\begin{figure}[h]
\centering
\includegraphics[width=0.4\textwidth]{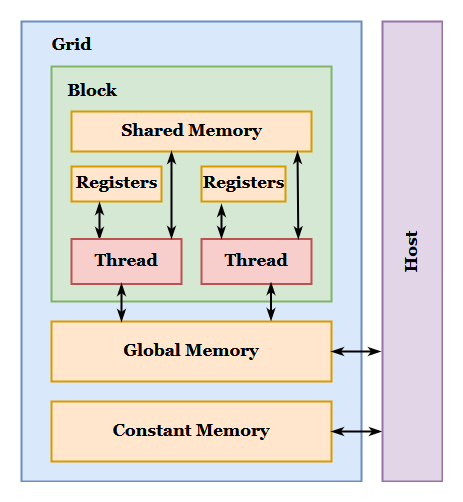}
\caption{Cuda memory architecture showing the relation between grid, block and threads and the corresponding proximity and connections of different kinds of memory elements}
\label{fig:cudarch}
\end{figure}

The CUDA application programming interface provides a way to structure our operations to run on Nvidia GPUs. The memory model in CUDA is divided into grids, blocks and threads which have access to specific kinds of memory and are all interfaced with the CPU, called a host, via a PCI bus as shown in Figure \ref{fig:cudarch}. Code execution can be segmented to run in grids and blocks, both of which can be molded to have one to three dimensions depending on the problem. Each block runs the kernel, a block of CUDA procedure, in individual threads. Threads within a block can share a portion of memory called shared memory, which has very low latency as it resides on the chip. Current GPUs can run 32 threads in a block, which is also called a warp, at a single time where same instructions of a kernel executes on all the threads but runs on different data. Optimizing for efficient allocation of warps could lead to better performance. Threads in a block can be synchronized with \_\_syncthreads() API call which syncs all the threads in a block at the code location where all of them execute \_\_syncthreads(). Concept of unified memory was introduced in CUDA 4.0 which does away with the manual process of memory copying from device to host and back. Now, we allocate a portion of memory that is shared between both device and host and GPU driver takes care of transfer of data when data is accessed from either device. We use cudaMallocManaged() to allocate shared memory and use cudaDeviceSynchronize() before accessing any device data from the host. 

\subsection{Sequential activation}

\begin{algorithm}
   \caption{Network segmentation algorithm}
   \label{alg:segment}
    \begin{algorithmic}[1]
      \Function{Segment\_network}{$R,IN,OP,CON$}
      \Comment{R=required, IN=inputs, OP=outputs, CON=connections}
        \State L $\gets$ []
        \State $s \gets IN$
        \While {True}
        
            \Comment{Candidate nodes for the next layer}
            \State c = \{ b \textbf{for} (a,b) \textbf{in} CON \textbf{if} a \textbf{in} s \textbf{and} b \textbf{not in} s\}
            
            \Comment{Used nodes whose entire input set is in s}
            \State $t \gets \{\}$
            \For {n \textbf{in} c}
            \State all $\gets$ (\textbf{for} a \textbf{in} s \textbf{for} (a,b) \textbf{in} CON \textbf{if} b = n)
            \If {n \textbf{in} R \textbf{and} all}
                \State t $\gets$ t  $U$ n
            \EndIf   
            \EndFor
            \If {t = $\phi$}
                \State \textbf{break}
            \EndIf 
            \State L $\gets$ L + t
            \State s $\gets$ s $U$ t
        \EndWhile
        \State \textbf{return} L
       \EndFunction
\end{algorithmic}
\end{algorithm}

For sequential activation of arbitrary neural network, we first perform pre processing on network structure to segment the network into sequential hierarchy of nodes, which we call layers of the ASNN. The algorithm to segment the network is given in Algorithm~\ref{alg:segment}. The function takes all nodes, input nodes,  output nodes and a structure containing all the network connections as input for processing. First, we find candidate nodes from the connections pool based on the nodes that have already been assigned to some layer. The candidates are those nodes for which incoming nodes all lie on the nodes that have already been assigned to a layer and all its outgoing nodes are not in the assigned set. From the candidate set, we only add those nodes to the next layer if their entire input set is contained in the nodes which are already assigned a layer value.

For sequential propagation of the neural network, All the nodes starting of the input layer is sequentially activated till the output layer from which the answer is obtained. The activation is calculated by going through all the incoming nodes and multiplying the connection weights with the node values and then summing them and then squashing them with the sigmoid function.

\subsection{Parallel GPU activation}

For single GPU activation of the neural network, we use the fact that all the nodes belonging to the same layer can be activated at once without compromising the output of the network in any way. For representation of the structure of network in the GPU, we use a custom data structure called CudaNode as depicted in Algorithm \ref{alg:structlisting}. Each CudaNode structure represents a single node of the neural network where each node contains a unique id, the number of incoming nodes, the integer array containing the node ids for the incoming connections and another float array for the corresponding weights for the incoming nodes. Then, we also have a Boolean that specifies if the node is a sensor i.e. takes external input for the network. The layer variable is set by the prepossessing of the neural network sequentially. The array of CudaNode structs are sorted in ascending order based on their layer number, where the input layer starts with value 0 and then climbs up to the last layer in the network. This is done to allow for better cache performance for unified memory in the GPU as input  and output nodes will be close to each other in the array after being sorted. 

\begin{algorithm}[t]
   \caption{Data structure to represent a single node in a network}
   \label{alg:structlisting}
    \begin{algorithmic}[1]
    \State \textbf{struct} CudaNode
        \State \hspace{\algorithmicindent} Integer: id 
        \Comment{Unique id for the node}
        \State \hspace{\algorithmicindent} Integer: layer 
        \Comment{Layer the node is in}
        \State \hspace{\algorithmicindent} Integer: numInNodes
        \Comment{No. of incoming nodes}
        \State \hspace{\algorithmicindent} Boolean: isSensor 
        \Comment{True if input node}
        \State \hspace{\algorithmicindent} Integer[]: inNodes 
        \Comment{Array of incoming Node ids}
        \State \hspace{\algorithmicindent} Float[]: inWeights 
        \Comment{Array of incoming Node weights}
    \State \textbf{end struct}
\end{algorithmic}
\end{algorithm}

CUDA kernel for GPU activation is described in Algorithm \ref{alg:cudakernel}.
The kernel for GPU activation takes the value for total number of layers in the network, another integer array containing number of nodes in each layer, the main CudaNodes array containing sorted node entities, and an input array of floats which contains values for input layer of the neural network. Size of the output float array is equal to the size of number of nodes in the network as each node writes its activation result to this array, which all other nodes will also be able to observe and write to. We have a variable $cl$ which determines the current layer that the kernel is processing and $sid$, the start id which holds the id of the first node of the layer being computed. We will already have spawned many threads which will correspond to one node of the layer being activated. In case the current node is a sensor, we just perform sigmoid activation on the input variable from input array corresponding to the current node id. If the current node is not a sensor, we sum the values from all the incoming nodes after weighting them with the connection weight then do the sigmoid activation on the resulting sum.

\begin{algorithm}
   \caption{CUDA kernel for calculating activation for ASNNs}
   \label{alg:cudakernel}
    \begin{algorithmic}[1]
        \State $\triangleright$ Integer: TL = Total layers in a network
        \State $\triangleright$ Integer[]: NNL = Number of nodes in layers
        \State $\triangleright$ CudaNode[]: n = Array of all the nodes
        \State $\triangleright$ Float[]: in = Input values for network
        \State $\triangleright$ Float[]: op = Array for output values
        \Function{cuda\_activation}{TL, NNL, n, in, op}
            \State Integer: cl $\gets$ 0
            \Comment{Current Layer}
            \State Integer: sid $\gets$ 0
            \Comment{Start id}
            \State Integer: id $\gets$ threadIdx.x
            \While {cl $<$ TL \textbf{and} id $<$ NNL$_{c1}$ }
                \State CudaNode: cn $\gets$ n$_{sid + id}$ 
                 \If {cn \textbf{is a} sensor}
                    \State op$_{cn.id}$ $\gets$ \textbf{call} sigmoid(in$_{cn.id}$)
                \Else
                    \State Float: sum $\gets$ 0
                     \For {i \textbf{from} 0 $\to$ cn.numInNodes - 1}
                        \State sum += cn.inWeights$_{i}$ * op$_{cn.inNodes[i]}$     
                    \EndFor
                \State op$_{cn.id} \gets$ \textbf{call} sigmoid(sum)
                \EndIf  
                \State \textbf{call} \_\_syncthreads()
                \State sid $\gets$ sid + NNL$_{c1}$
                \State cl $\gets$ cl + 1
            \EndWhile
       \EndFunction
\end{algorithmic}
\end{algorithm}

After one activation of one node is completed, we call \_\_syncthreads() in the kernel to wait for all other nodes on the layer to finish computing. After synchronization, we increase the current layer variable to denote that we have progressed one layer of the neural network and increase start id by the number of nodes which were present in the completed layer. We compare current layer variable against total number of layers of the neural network and exit out of the loop if current layer is greater then the total layers, which signifies that the network has completed activation. After completing activation, we make sure to call cudaDeviceSynchronize() function before we read in the answers from the host to give the host enough time to copy the results from the device memory to host memory. Then, we read in the values from output array which has the final activations of output nodes in the network.

\subsection{Experimental setup}

We used the CUBIX machine in our department for running all our experiments, which had the following configuration:

\begin{itemize}
\item Two 6 core Intel Xeon CPUs (E5-2620 0 @ 2.00GHz)
\item CPU caches of L1: 32K, L2: 256K, L3: 15360K
\item 64 GB of RAM
\item 8 GTX 1080 with 8GB DRAM each
\end{itemize}

For all the results that follow, experiments were run 10 times and averaged for GPU activation timings and run 5 times and averaged for sequential activation timings.

\section{Results and discussion} \label{results}

\subsection{Sequential results}

\begin{figure}[h]
\centering
\includegraphics[width=\linewidth]{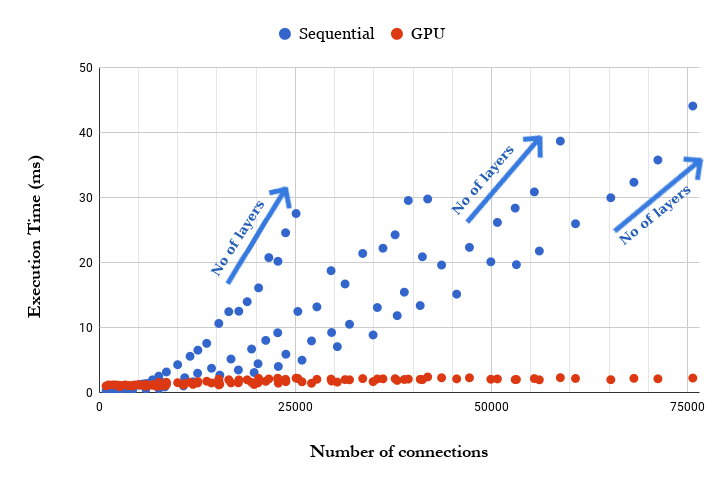}
\vspace{-20pt}
\caption{Increase in execution time with respect to number of connections in a network for Sequential(blue) and GPU(red) approaches}
\label{fig:seqvsgpu}
\end{figure}

From Figure \ref{fig:seqvsgpu} we can see that the execution of networks take linear amount of time with respect to the number of connections in the network. Increase in number of layers also correspond with the increase in execution time. Thus, with large number of connections and many layers, the execution time drastically increases. It is also possible for a network with same number of connection to have different execution time based on number of layers, as even with same number of connections, network with deeper layers take longer to execute.

\subsection{Single GPU result}

\begin{figure}[h]
\centering
\includegraphics[width=\linewidth]{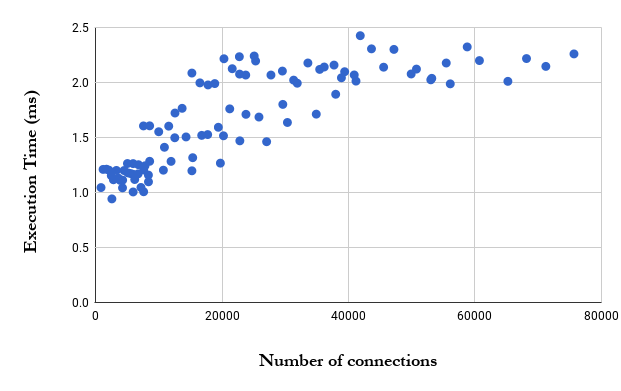}
\vspace{-20pt}
\caption{Increase in execution time with respect to number of connections in a network using GPU implementation}
\label{fig:gpu}
\end{figure}

From Figure \ref{fig:gpu} we see that there is a general upward trend for the execution time with respect to the number of connections. But it should be noted that the slope is very flat compared to the sequential execution graph. The minimum is at 1ms and the maximum at 2.5ms.

\subsection{Comparison and speedup}

\begin{figure}[h]
\centering
\includegraphics[width=\linewidth]{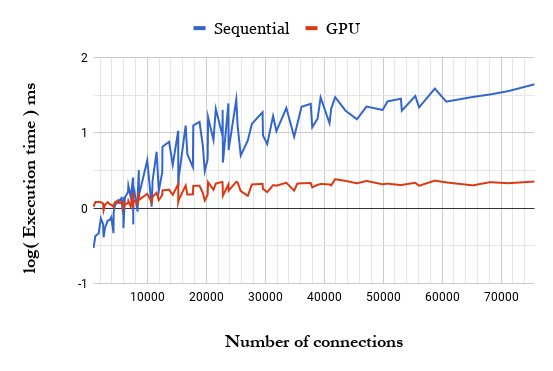}
\vspace{-20pt}
\caption{Log of execution time with respect to number of connections in network, showing the comparison of execution times for Sequential and GPU approaches}
\label{fig:logseqvsgpu}
\end{figure}

From Figure \ref{fig:seqvsgpu} we notice that compared to sequential execution, the GPU execution time lies flat and skims the x-axis. We can also compare the log of execution time from Figure \ref{fig:logseqvsgpu} where we see that the sequential execution time is very low for smaller networks but grows log linear with the number of connections. The log graph of execution time for sequential method has a steep slope at start, and still  has positive slope after 30000 connections, while the GPU execution curve is flat with the x axis after 30000 connections. Thus, we can be certain that the GPU approach will scale well with increase in number of connections.

\begin{figure}[h]
\centering
\includegraphics[width=\linewidth]{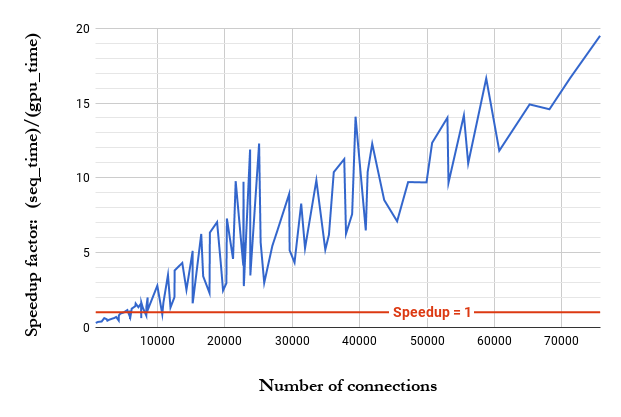}
\vspace{-20pt}
\caption{Increase in speedup factor with increase in number of connections in a network showing the linear increase in speedup. Red line shows where speedup crosses the factor of 1}
\label{fig:speedup}
\end{figure}

For speedup, we can see from the Figure \ref{fig:speedup} that initially, the speedup factor of sequential timings compared with GPU timings are lower than one which means that the GPU approach is slower than the sequential approach for very small networks. This is predictable as overhead of copying to and from the device negates any speedup from the computation happening at the device. Till 8000 connections, speedup from GPU is similar to the the sequential implementation, but after 8000 connections, speedup shows linear increase with number of connections. The jagged structure of the line is due to the variation in number of layers possible for a same connection count i.e. low depth networks can be computed quickly compared to deeper networks. From the graph, we can see that the speedup factor is fairly high for low depth networks too and is linearly increasing with increase in number of connections. This results signify that we will have more gains the larger our network gets. If we take a network with 70,000 connections, we can get up to 15 times speedup, which will have huge gains given that the networks are evaluated for thousands of iterations for any given problem.

\section{Conclusion and Future Work} \label{conclusion}

Our research focused on finding effective ways of accelerating arbitrary structured neural networks. We were able to show that: by pre-processing the network to segment it into dependent layers and then using CUDA threads to execute all the nodes in the same layer at the same time, we can get speedup that increases with the size of the connections in the neural network. From our experimentation, we have shown linear speedup increase for our GPU implementation compared to our sequential implementation.

We can further improve on this work by extending the approach to incorporate grid wide thread locking to synchronize threads in a grid group which will significantly increase the number of nodes which can be processed simultaneously. The natural extension of this work is, to find ways to perform network segmentation in GPU itself; which will also have significant impact on the overall efficiency of current approach. Our approach of using GPUs to accelerate arbitrary neural networks can also be used for other domains of research, as networks found in nature generally have non uniform structure, so our research can be incorporated for their study and simulation. One prominent example is of biological brains which have arbitrary structure with huge number of nodes and connections. We could simulate and study such large networks if we can extend the processing capacity by coordinating multiple GPUs to compute a single network.

\ifCLASSOPTIONcaptionsoff
  \newpage
\fi

\printbibliography 

\end{document}